\documentclass[twocolumn,twocolappendix]{aastex701}
\usepackage{CJKutf8}
\usepackage{booktabs}
\usepackage{threeparttable}
\usepackage{siunitx}

\usepackage{placeins}

\begin{document}

\title{WFST0533buha (Bi'an): A Fast Blue Ultraluminous Transient Candidate with Unusual Blue/NUV Colors}

\author[0009-0009-5238-2059]{Zelin Xu}
\affiliation{Department of Astronomy, University of Science and Technology of China, Hefei 230026, China}
\affiliation{School of Astronomy and Space Sciences, University of Science and Technology of China, Hefei, 230026, China}
\email{sa21022025@mail.ustc.edu.cn}

\author[0000-0002-9092-0593]{Ji-an Jiang}
\affiliation{Department of Astronomy, University of Science and Technology of China, Hefei 230026, China}
\affiliation{School of Astronomy and Space Sciences, University of Science and Technology of China, Hefei, 230026, China}
\affiliation{National Astronomical Observatory of Japan, 2-21-1 Osawa, Mitaka, Tokyo 181-8588, Japan}
\textcolor{blue}{\email{jian.jiang@ustc.edu.cn}}

\author[0009-0004-0078-8982]{Ziqing Jia}
\affiliation{Department of Astronomy, University of Science and Technology of China, Hefei 230026, China}
\affiliation{School of Astronomy and Space Sciences, University of Science and Technology of China, Hefei, 230026, China}
\email{zqjia@mail.ustc.edu.cn}

\author[0000-0003-2611-7269]{Keiichi Maeda}
\affiliation{Department of Astronomy, Kyoto University, Kitashirakawa-Oiwake-cho, Sakyo-ku, Kyoto 606-8502, Japan}
\email{keiichi.maeda@kusastro.kyoto-u.ac.jp}

\author{Dezheng Meng}
\affiliation{Department of Astronomy, University of Science and Technology of China, Hefei 230026, China}
\affiliation{School of Astronomy and Space Sciences, University of Science and Technology of China, Hefei, 230026, China}
\email{dezhengmeng@mail.ustc.edu.cn}

\author[0000-0002-2242-1514]{Zhengyan Liu}
\affiliation{Department of Astronomy, University of Science and Technology of China, Hefei 230026, China}
\email{ustclzy@mail.ustc.edu.cn}

\author[0009-0003-9474-8457]{Weiyu Wu}
\affiliation{Department of Astronomy, University of Science and Technology of China, Hefei 230026, China}
\affiliation{School of Astronomy and Space Sciences, University of Science and Technology of China, Hefei, 230026, China}
\email{wwyyy@mail.ustc.edu.cn} 

\author{Junhan Zhao}
\affiliation{Department of Astronomy, University of Science and Technology of China, Hefei 230026, China}
\affiliation{School of Astronomy and Space Sciences, University of Science and Technology of China, Hefei, 230026, China}
\email{zjunhan@mail.ustc.edu.cn}

\author{Hanindyo Kuncarayakti}
\affiliation{Tuorla Observatory, Department of Physics and Astronomy, FI-20014 University of Turku, Finland}
\affiliation{Finnish Centre for Astronomy with ESO (FINCA), FI-20014 University of Turku, Finland}
\email{kuncarayakti@gmail.com}

\author{Lluís Galbany}
\affiliation{Institute of Space Sciences (ICE-CSIC), Campus UAB, Carrer de Can Magrans, s/n, E-08193 Barcelona, Spain}
\affiliation{Institut d'Estudis Espacials de Catalunya (IEEC), 08860 Castelldefels (Barcelona), Spain}
\email{lluisgalbany@gmail.com}

\author{Saurabh W. Jha}
\affiliation{Department of Physics and Astronomy, Rutgers, The State University of New Jersey, 136 Frelinghuysen Road, Piscataway, New Jersey 08854, USA}
\email{saurabh@physics.rutgers.edu}

\author{\v{Z}eljko Ivezi\'{c}}
\affiliation{Department of Astronomy, University of Washington, Box 351580, Seattle, Washington 98195-1580, USA}
\email{ivezic@uw.edu}

\author{Peter Yoachim}
\affiliation{Department of Astronomy, University of Washington, Box 351580, Seattle, Washington 98195-1580, USA}
\email{yoachim@uw.edu}

\author{Andrew J. Connolly}
\affiliation{Department of Astronomy, University of Washington, Box 351580, Seattle, Washington 98195-1580, USA}
\email{ajc@astro.washington.edu}

\author[0000-0003-4341-0029]{Tao An}
\affiliation{Department of Astronomy, University of Science and Technology of China, Hefei 230026, China}
\email{antao2008@ustc.edu.cn}

\author[0009-0009-2057-7564]{Ming Chen}
\affiliation{Department of Astronomy, University of Science and Technology of China, Hefei 230026, China}
\affiliation{School of Astronomy and Space Sciences, University of Science and Technology of China, Hefei, 230026, China}
\email{mingchen26@mail.ustc.edu.cn}

\author[0000-0002-7835-8585]{Zigao Dai}
\affiliation{Department of Astronomy, University of Science and Technology of China, Hefei 230026, China}
\affiliation{School of Astronomy and Space Sciences, University of Science and Technology of China, Hefei, 230026, China}
\email{daizg@ustc.edu.cn}

\author[0000-0002-7660-2273]{Xu Kong}
\affiliation{Department of Astronomy, University of Science and Technology of China, Hefei 230026, China}
\affiliation{School of Astronomy and Space Sciences, University of Science and Technology of China, Hefei, 230026, China}
\affiliation{Institute of Deep Space Sciences, Deep Space Exploration Laboratory, Hefei 230026, China}
\email{xkong@ustc.edu.cn}

\author[0009-0002-8721-6201]{Tiancheng Lv}
\affiliation{Department of Astronomy, University of Science and Technology of China, Hefei 230026, China}
\affiliation{School of Astronomy and Space Sciences, University of Science and Technology of China, Hefei, 230026, China}
\email{tclv@mail.ustc.edu.cn}

\author[0000-0001-8253-6850]{Masaomi Tanaka}
\affiliation{Astronomical Institute, Tohoku University, Aoba, Sendai 980-8578, Japan}
\email{masaomi.tanaka@astr.tohoku.ac.jp}

\author[0000-0001-8537-3153]{Nozomu Tominaga}
\affiliation{National Astronomical Observatory of Japan, 2-21-1 Osawa, Mitaka, Tokyo 181-8588, Japan}
\affiliation{SOKENDAI (The Graduate University for Advanced Studies), Mitaka, Tokyo, 181-8588, Japan}
\email{nozomu.tominaga@nao.ac.jp}

\author{Ailing Wang}
\affiliation{Yunnan Observatories, Chinese Academy of Sciences, Kunming 650216, China}
\email{wangailing@ynao.ac.cn}

\author[0000-0002-7152-3621]{Ning Jiang}
\affiliation{Department of Astronomy, University of Science and Technology of China, Hefei 230026, China}
\affiliation{School of Astronomy and Space Sciences, University of Science and Technology of China, Hefei, 230026, China}
\email{jnacf@ustc.edu.cn}

\author[0000-0002-1935-8104]{Yongquan Xue}
\affiliation{Department of Astronomy, University of Science and Technology of China, Hefei 230026, China}
\affiliation{School of Astronomy and Space Sciences, University of Science and Technology of China, Hefei, 230026, China}
\email{xuey@ustc.edu.cn}

\author[0000-0002-1330-2329]{Wen Zhao}
\affiliation{Department of Astronomy, University of Science and Technology of China, Hefei 230026, China}
\affiliation{School of Astronomy and Space Sciences, University of Science and Technology of China, Hefei, 230026, China}
\email{wzhao7@ustc.edu.cn}

\author{Qingfeng Zhu}
\affiliation{Department of Astronomy, University of Science and Technology of China, Hefei 230026, China}
\affiliation{School of Astronomy and Space Sciences, University of Science and Technology of China, Hefei, 230026, China}
\email{zhuqf@ustc.edu.cn}

\author[0000-0003-4200-4432]{Lulu Fan}
\affiliation{Department of Astronomy, University of Science and Technology of China, Hefei 230026, China}
\affiliation{School of Astronomy and Space Sciences, University of Science and Technology of China, Hefei, 230026, China}
\affiliation{Institute of Deep Space Sciences, Deep Space Exploration Laboratory, Hefei 230026, China}
\affiliation{College of Physics, Guizhou University, Guiyang 550025, China}
\email{llfan@ustc.edu.cn}

\author{Feng Li}
\affiliation{State Key Laboratory of Particle Detection and Electronics, University of Science and Technology of China, Hefei 230026, China}
\email{phonelee@ustc.edu.cn}

\author{Ming Liang}
\affiliation{National Optical Astronomy Observatory (NSF’s National Optical-Infrared Astronomy Research Laboratory) 950 N Cherry Ave. Tucson Arizona 85726, USA}
\email{liangming@gmail.com}

\author{Binyang Liu}
\affiliation{Purple Mountain Observatory, Chinese Academy of Sciences, Nanjing 210023, China}
\email{byliu@pmo.ac.cn}

\author[0000-0002-3105-3821]{Zhen Wan}
\affiliation{Department of Astronomy, University of Science and Technology of China, Hefei 230026, China}
\email{zhen_wan@ustc.edu.cn}

\author[0000-0002-4372-0759]{Hairen Wang}
\affiliation{Purple Mountain Observatory, Chinese Academy of Sciences, Nanjing 210023, China}
\email{hairenwang@pmo.ac.cn}

\author[0000-0003-1617-2002]{Jian Wang}
\affiliation{State Key Laboratory of Particle Detection and Electronics, University of Science and Technology of China, Hefei 230026, China}
\affiliation{Institute of Deep Space Sciences, Deep Space Exploration Laboratory, Hefei 230026, China}
\email{wangjian@ustc.edu.cn}

\author{Ji Yang}
\affiliation{Purple Mountain Observatory, Chinese Academy of Sciences, Nanjing 210023, China}
\email{jiyang@pmo.ac.cn}

\author{Hongfei Zhang}
\affiliation{State Key Laboratory of Particle Detection and Electronics, University of Science and Technology of China, Hefei 230026, China}
\email{nghong@ustc.edu.cn}

\author[0000-0003-3728-9912]{Xianzhong Zheng}
\affiliation{Tsung-Dao Lee Institute and Key Laboratory for Particle Physics, Astrophysics and Cosmology, Ministry of Education, Shanghai Jiao Tong University, Shanghai, 201210, China}
\email{xzzheng@sjtu.edu.cn}

\correspondingauthor{Ji-an Jiang}
\email{jian.jiang@ustc.edu.cn}

\begin{abstract}
We present WFST0533buha (Bi'an, ``\begin{CJK*}{UTF8}{gbsn}狴犴\end{CJK*}''), a fast blue ultraluminous transient (FBUT) candidate discovered by the Wide Field Survey Telescope Deep High-cadence $ugr$-band Survey. The source brightened by approximately 2.5 mag in $u$ within 3 days and has a half-maximum timescale ($t_{1/2}$) shorter than 6 days. Two spectra show a blue, featureless continuum but do not provide a secure redshift. Einstein Probe observations yield only X-ray upper limits, while a faint radio counterpart was detected at 15 GHz with the VLA. The early $u$-band detection and intensive multiband coverage of Bi'an demonstrate the value of deep, high-cadence blue-band observations for discovering and characterizing FBUTs at early phases. Notably, Bi'an shows redder blue/NUV colors than AT 2018cow and AT 2024wpp, suggesting possible diversity in the short-wavelength properties of FBUTs.\end{abstract}

\section{Introduction}
Wide-field time-domain surveys have opened up a new region of transient parameter space, revealing populations of events spanning a broad range of luminosities and timescales. Among these, fast blue optical transients (FBOTs) have emerged as a distinct class. Empirically, they are often identified by blue colors and short durations, with light curves remaining above half of their peak luminosity (t$_{1/2}$) for fewer than 12 days \citep{drout_14,tanaka_16,pursiainen_18,Ho_2023}.

Among rapidly evolving blue transients, fast blue ultraluminous transients (FBUTs), also commonly referred to as luminous fast blue optical transients (LFBOTs), represent an observationally extreme population characterized by rapid rises and declines, high peak optical luminosities($M_g \lesssim -20\,\mathrm{mag}$), and blue continua at early times \citep{Margutti_2019,perley_2019,jiang_2022,Metzger_2022}. In several well-observed cases, luminous radio and/or X-ray emission has been detected, with some events showing strong X-ray variability on timescales of minutes to days \citep{perley_2019,Margutti_2019,Coppejans_2020,Bright_2022,Yao_2022,Chrimes_2024a,Nayana_2025}. Such properties—particularly the combination of short timescales and sustained high luminosities—are difficult to reconcile with ordinary $^{56}\text{Ni}$-powered supernova models. Instead, they point toward a small effective ejecta mass coupled with an additional sustained power source and dense surrounding material \citep{Margutti_2019,Bright_2022,Yao_2022,Chrimes_2024a,Nayana_2025}. Since the landmark discovery of the prototype AT~2018cow \citep{Prentice_2018,Ho_2019,perley_2019,Margutti_2019}, a growing number of similar events have been identified, demonstrating that these extreme properties are not unique to AT~2018cow and pose a recurring challenge to our understanding of stellar death \citep{Ho_Lu_2026}.

Despite the expansion of the observational census, the physical origin of FBUTs remains unsettled. Proposed interpretations include the collapse or failed explosion of a massive star, with core collapse forming a black hole and fallback accretion supplying a sustained power source after the initial explosion \citep{perley_2019,Margutti_2019,Quataert_2019,Guti_2024}; 
tidal-disruption-like events involving an accreting compact object, in which the disruption of a star and the resulting accretion flow can naturally produce luminous and variable high-energy emission \citep{perley_2019,Kuin_2019,Uno_2020,Guti_2024,Ho_2023a}; and merger-driven channels involving a stellar-mass black hole or neutron star and a Wolf-Rayet (WR) star, in which tidal stripping, hyperaccretion, and disk-driven outflows may account for the low ejecta mass, dense circumstellar material (CSM), and aspherical geometry inferred for the class \citep{Metzger_2022,Klencki_2026}. Other proposed powering and emission mechanisms include magnetar-powered engines \citep{Prentice_2018,Li_2024}, strong CSM-ejecta interaction, jet/cocoon models \citep{Gottlieb_2022}, and shock-cooling scenarios \citep{Ofek_2010,pursiainen_18}, which may help explain the rapid blue emission and the observed multiwavelength diversity. Many of these mechanisms are compatible with the progenitor scenarios discussed above.

Recent population studies have begun to reveal both commonality and diversity within the class. Several FBUTs are found in actively star-forming, relatively low-mass host galaxies, broadly consistent with the environments expected for young massive-star progenitors, yet other events show larger offsets or less direct links to ongoing star formation \citep{Nugent_2026,Sevilla_2026}. These results raise the possibility that FBUTs may arise from more than one physical channel, or that a single underlying mechanism can operate across a wider range of progenitor environments than previously assumed. In either case, each new well-observed event provides an important opportunity to test whether the defining properties of the class are truly universal. In this Letter, we report the discovery of a new FBUT candidate, WFST0533buha (Bi'an, ``\begin{CJK*}{UTF8}{gbsn}狴犴\end{CJK*}"\footnote{Bi'an is one of the nine sons of the Chinese dragon, a tiger-like mythical creature known for its fierce sense of justice, whose image traditionally adorns tribunal gates and prison doors.}; AT~2025epd \citep{2025TNS}), discovered with the Wide Field Survey Telescope (WFST), and compare its observational properties with those of the growing FBUT population.

This Letter is organized as follows. Section \ref{sec:obs} presents the observations and data reduction. In Section \ref{sec:an}, we analyze the main observational properties of WFST0533buha. Section \ref{sec:disc} discusses the implications of these observations for the FBUT population and possible progenitor scenarios. Throughout this Letter, we adopt a flat $\Lambda$CDM cosmology with $H_{0}=70\ \mathrm{km\ s^{-1}\ Mpc^{-1}}$ and $\Omega_{\mathrm{m}}=0.3$. Unless otherwise specified, all magnitudes are in the AB system.

\section{Observations and Data Reduction} \label{sec:obs}

The Wide Field Survey Telescope (WFST) is jointly built by the University of Science and Technology of China and the Purple Mountain Observatory. It is characterized by a 2.5-meter primary mirror and 9 $\times$ 9K $\times$ 9K mosaic CCDs with high sensitivity in blue wavelength. As the largest time-domain survey telescope in the northern hemisphere, the field of view of the WFST reaches 6.5 square degrees \citep{2023SCPMA..6609512W}.

The Deep High-cadence \textit{ugr}-band Survey (DH\textit{ugr}) is one of the major time-domain programs conducted by WFST, designed to repeatedly monitor selected sky regions with high cadence and multi-band coverage (Jiang et al. in prep). The images typically reach $5\sigma$ limiting magnitudes of approximately $22.9$, $23.3$, and $23.1$ mag in the $u$, $g$, and $r$ bands during photometric nights, respectively, with each field revisited every day. By combining relatively deep limiting magnitudes with dense temporal sampling, together with early-time $u$-band information, the survey is particularly well suited for the discovery and follow-up of fast blue transients, including FBOTs and FBUTs.

\subsection{Discovery of WFST0533buha} \label{subsec:discovery}

Bi'an was discovered on 3 March 2025, during the WFST DH\textit{ugr}. Its coordinates are $\alpha = 161.144035$, $\delta = +4.470170$ (J2000). Subsequent $u$-band photometry shows a brightening of $2.45\pm0.11$ mag between MJD 60737.748 and 60740.743, establishing a rapid rise over approximately 3 observer-frame days.

Following the discovery of its rapid rise, Bi'an was intensively monitored at optical wavelengths and targeted with additional spectroscopic, X-ray, and radio observations. The resulting multiwavelength data provide dense coverage of the early optical evolution together with constraints on its high-energy and radio emission.

The redshift of Bi'an remains unconstrained by the available observations. For convenience of presentation, we adopt $z=0.25$ as an illustrative value when deriving its rest-frame and distance-dependent properties throughout this work. The dependence of these quantities on redshift is examined in Section~4.1 and illustrated in Appendix~C.

\subsection{Optical Photometry} \label{subsec:optical}
Optical photometry of Bi'an in the \textit{ugr}-band was obtained by WFST from Modified Julian Dates (MJD) 60729 to 60764. The optical images are reduced using the WFST pipeline, which is based on the LSST pipeline \citep{Axelrod_2010,Bosch_2019,Ivezi_2019}. We performed standard image reduction including bias and flat correction, as well as astrometric calibration against the Gaia catalog and photometric calibration against the SDSS catalog (for the \textit{u}-band) or the PS1 catalog (for the \textit{gr}-band) with the WFST pipeline \citep{Cai_2025}. Image subtraction was performed using coadded template images taken in Nov. 2024 using the SFFT algorithm \citep{2022ApJ...936..157H}.

\begin{figure}[htbp]
	\includegraphics[width=\columnwidth]{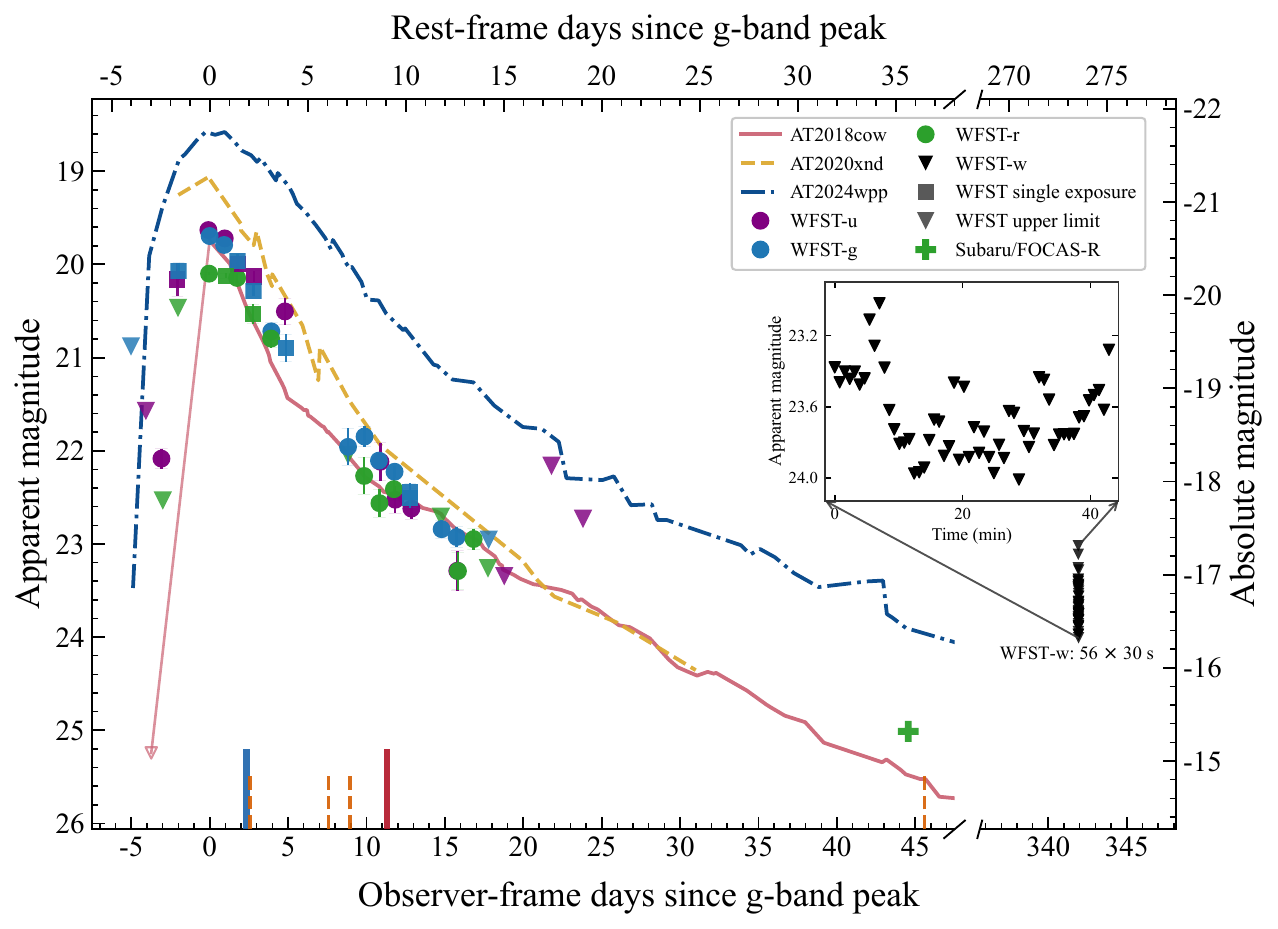}
	\caption{Multiband light curves of Bi'an at the fiducial redshift $z=0.25$, compared with the $g$-band light curves of AT~2018cow \citep{Prentice_2018,perley_2019}, AT~2020xnd \citep{Perley2021}, and AT~2024wpp \citep{Nayana_2025,LeBaron_2026}. $5\sigma$ nondetection limits are shown as downward triangles. The Galactic extinction has been corrected, assuming an extinction-to-reddening ratio of $R_V=3.1$. The colored solid bars and orange dashed bars along the bottom denote epochs of spectroscopy and EP/FXT X-ray observations, respectively. The right side shows the late-time WFST $w$-band intranight nondetection limits, with an inset highlighting the 56 consecutive 30 s exposures.}
	\label{fig:lc}
\end{figure}

When two exposures in the same band were obtained close in time during a night, we combined them before performing image subtraction and photometry. Otherwise, we used individual exposures. Figure \ref{fig:lc} presents the multi-band light curve of Bi'an. All WFST photometry was performed via PSF photometry.

In addition, we obtained Subaru/FOCAS coadded image on MJD 60785 in \textit{R} band. We performed PSF photometry directly on the coadded Subaru/FOCAS image, without host subtraction. In the total 1960 second coadded image, there is a clear detection with a magnitude of $25.15 \pm 0.18$ in the Vega system.

We corrected the data for Galactic extinction, adopting $E(B-V)=0.04$ mag and $R_V=3.1$ \citep{Schlegel_1998,Schlafly_2011}. No correction for host-galaxy extinction was applied because the available data do not provide an independent constraint on it. The corrected optical photometric results are listed in Table \ref{tab:buha_photometry}.

\begin{deluxetable}{l@{\hspace{0.8em}}c@{\hspace{0.8em}}c@{\hspace{0.9em}}c}
\tabletypesize{\scriptsize}
\tablewidth{0pt}
\tablecaption{Optical photometry of Bi'an corrected for Galactic extinction. Upper limits are in 5$\sigma$.\label{tab:buha_photometry}}
\tablehead{
\colhead{MJD} & \colhead{Filter} & \colhead{Magnitude} & \colhead{Instrument} \\
\colhead{(day)} & \colhead{} & \colhead{(mag)} & \colhead{}
}
\startdata
60735.806 & $g$ & $>20.97$ & WFST \\
60736.755 & $u$ & $>21.66$ & WFST \\
60736.798 & $g$ & $>21.65$ & WFST \\
60737.748 & $u$ & 22.08 $\pm$ 0.11 & WFST \\
60737.833 & $r$ & $>22.62$ & WFST \\
60738.764\tablenotemark{a} & $u$ & 20.16 $\pm$ 0.18 & WFST \\
60738.805 & $r$ & $>20.56$ & WFST \\
60738.842\tablenotemark{a} & $g$ & 20.07 $\pm$ 0.09 & WFST \\
60740.742 & $u$ & 19.63 $\pm$ 0.01 & WFST \\
60740.785 & $r$ & 20.10 $\pm$ 0.01 & WFST \\
60740.821 & $g$ & 19.70 $\pm$ 0.01 & WFST \\
60741.745 & $g$ & 19.79 $\pm$ 0.01 & WFST \\
60741.785 & $u$ & 19.72 $\pm$ 0.01 & WFST \\
60741.867\tablenotemark{a} & $r$ & 20.12 $\pm$ 0.03 & WFST \\
60742.573 & $r$ & 20.14 $\pm$ 0.05 & WFST \\
60742.593\tablenotemark{a} & $g$ & 19.97 $\pm$ 0.08 & WFST \\
60742.630\tablenotemark{a} & $u$ & 20.00 $\pm$ 0.07 & WFST \\
60743.589\tablenotemark{a} & $r$ & 20.53 $\pm$ 0.11 & WFST \\
60743.625\tablenotemark{a} & $g$ & 20.28 $\pm$ 0.06 & WFST \\
60743.662\tablenotemark{a} & $u$ & 20.13 $\pm$ 0.06 & WFST \\
60744.740 & $r$ & 20.80 $\pm$ 0.10 & WFST \\
60744.754 & $g$ & 20.72 $\pm$ 0.09 & WFST \\
60745.621 & $u$ & 20.50 $\pm$ 0.14 & WFST \\
60745.674\tablenotemark{a} & $g$ & 20.90 $\pm$ 0.15 & WFST \\
60749.645 & $r$ & $>22.14$ & WFST \\
60749.647 & $g$ & 21.96 $\pm$ 0.20 & WFST \\
60750.671 & $r$ & 22.27 $\pm$ 0.20 & WFST \\
60750.701 & $g$ & 21.85 $\pm$ 0.11 & WFST \\
60751.618 & $g$ & 22.11 $\pm$ 0.08 & WFST \\
60751.649 & $r$ & 22.56 $\pm$ 0.15 & WFST \\
60751.713 & $u$ & 22.12 $\pm$ 0.20 & WFST \\
60752.572 & $r$ & 22.41 $\pm$ 0.12 & WFST \\
60752.613 & $g$ & 22.23 $\pm$ 0.06 & WFST \\
60752.654 & $u$ & 22.52 $\pm$ 0.15 & WFST \\
60753.607\tablenotemark{a} & $g$ & 22.44 $\pm$ 0.09 & WFST \\
60753.608\tablenotemark{a} & $g$ & 22.51 $\pm$ 0.10 & WFST \\
60753.685 & $u$ & 22.62 $\pm$ 0.11 & WFST \\
60755.571 & $r$ & $>22.80$ & WFST \\
60755.616 & $g$ & 22.84 $\pm$ 0.10 & WFST \\
60755.657 & $u$ & $>23.00$ & WFST \\
60756.574 & $g$ & 22.93 $\pm$ 0.11 & WFST \\
60756.614 & $u$ & 23.29 $\pm$ 0.21 & WFST \\
60756.655 & $r$ & 23.29 $\pm$ 0.20 & WFST \\
60757.655 & $r$ & 22.95 $\pm$ 0.11 & WFST \\
60758.573 & $r$ & $>23.35$ & WFST \\
60758.613 & $g$ & $>23.05$ & WFST \\
60758.654 & $u$ & $>23.12$ & WFST \\
60759.608 & $u$ & $>23.44$ & WFST \\
60759.648 & $g$ & $>21.89$ & WFST \\
60760.612 & $u$ & $>22.28$ & WFST \\
60762.623 & $u$ & $>22.25$ & WFST \\
60763.704 & $u$ & $>22.52$ & WFST \\
60764.630 & $u$ & $>22.82$ & WFST \\
60785.370 & $R$ & 25.06 $\pm$ 0.18 & Subaru/FOCAS \\
\enddata
\tablecomments{Magnitudes have been corrected for Galactic extinction using $A_u=0.163$, $A_g=0.127$, and $A_r=0.088$ mag. The WFST magnitudes are in the AB system, while the Subaru/FOCAS $R$-band magnitude is in the Vega system.}
\tablenotetext{a}Measurements from individual exposures. All other measurements are from coadded exposures.
\end{deluxetable}

\subsection{Spectroscopy} \label{subsec:spec}
\begin{figure}
    \centering
    \includegraphics[width=\linewidth]{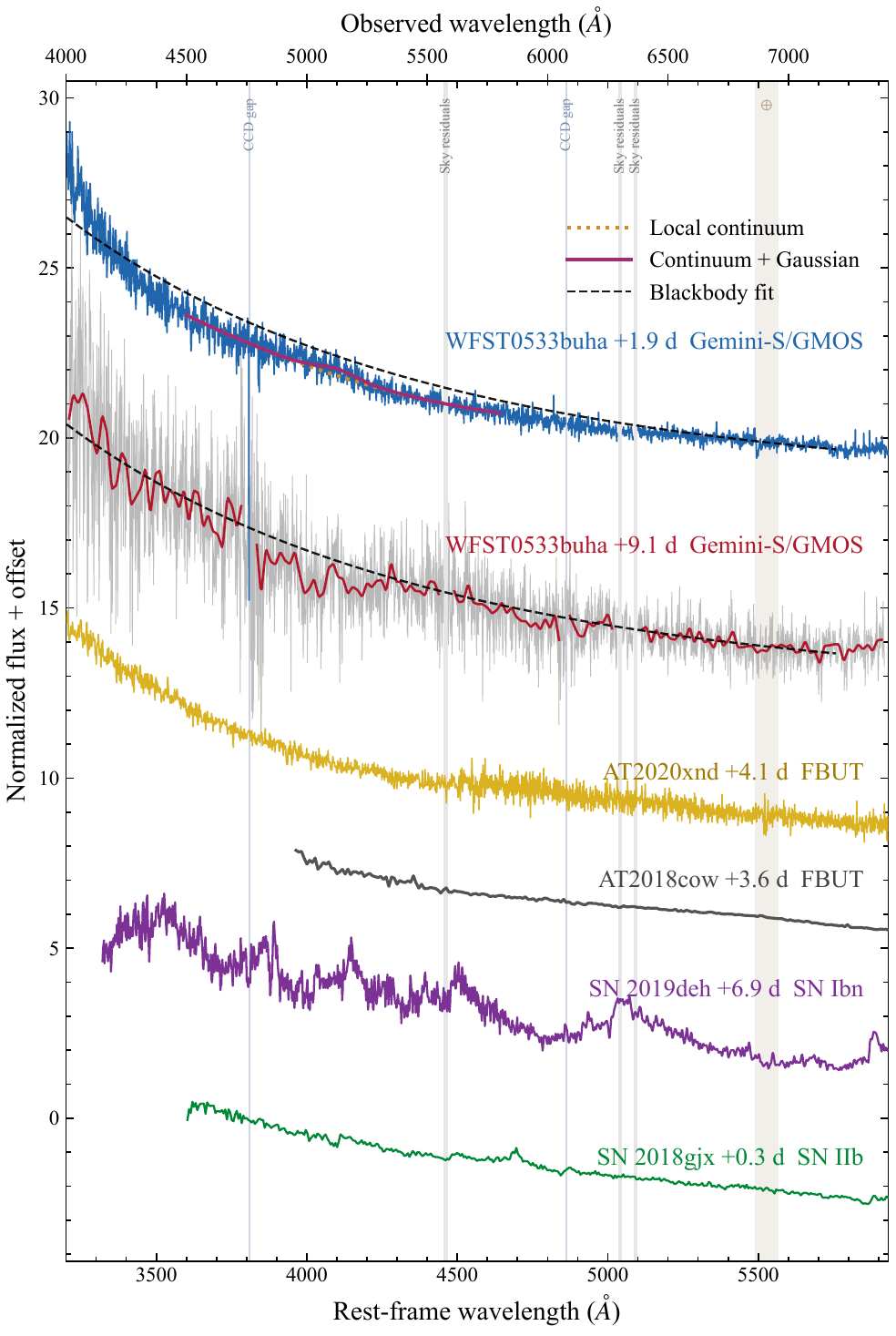}
    \caption{Spectra of Bi'an at approximately 2 and 11 observer-frame days after maximum, compared with AT~2018cow \citep{Prentice_2018}, AT~2020xnd \citep{Perley2021}, SN~2019deh \citep{Pellegrino_2022}, and SN~2018gjx \citep{Prentice_2020}. The labels indicate rest-frame phases assuming $z=0.25$ for Bi'an. The second spectrum is shown in 20 \AA\ bins (red), with the unbinned spectrum in grey. Black dashed curves show blackbody fits; orange dotted and purple solid curves show the local continuum and continuum-plus-Gaussian fit, respectively. Shaded bands mark detector gaps, sky residuals, and telluric absorption.}
    \label{fig:spec}
\end{figure}

We obtained two epochs of optical spectroscopy of Bi'an with the Gemini Multi-Object Spectrograph on the Gemini South telescope (GMOS-S; \citealt{Hook_2004}) on 9 March and 18 March 2025. At each epoch, four exposures were obtained with different dithering positions in both the spatial and dispersion directions, using the B480 grating and a 1.0 arcsec slit. Individual exposure times were 450 s on 9 March and 1200 s on 18 March, corresponding to total integration times of 1800 s and 4800 s, respectively. The data were processed with the Gemini DRAGONS reduction pipeline following standard procedures, including bias subtraction, flat-field correction, wavelength calibration, sky subtraction, spectral extraction, and flux calibration \citep{Labrie2019}. Pixels flagged by the DRAGONS data-quality arrays were excluded from the analysis. The two spectra are shown in Figure~\ref{fig:spec}, both dominated by blue, featureless continua.

\subsection{VLA observations}

We observed Bi'an with the Karl G. Jansky Very Large Array (VLA) at multiple frequencies between 2025 March 29 and 2025 July 1. The observations were carried out in the S, C, X, and Ku bands, with central frequencies of 3, 6, 10, and 15 GHz, respectively. The first epoch on 2025 March 29 provided the broadest frequency coverage, and later epochs on 2025 May 1 and 2025 July 1 were obtained at X and Ku bands. The synthesized beam sizes ranged from $23\arcsec \times 18\arcsec$ at 3 GHz to $1.5\arcsec \times 1.2\arcsec$ at 15 GHz, reflecting the different observing frequencies and array configurations. A summary of the observations, including the beam size, position angle, image rms noise, and peak flux density measurements, is given in Table~\ref{tab:obs_summary}.

The VLA data were calibrated using the Common Astronomy Software Applications package (\textsc{CASA}). Standard VLA calibration procedures were applied, including initial flagging of radio-frequency interference, delay calibration, bandpass calibration, complex gain calibration, and absolute flux-density calibration \citep{CASA_2022}. The calibrated visibilities were inspected, and additional flagging was applied where necessary. Imaging was performed with the \texttt{tclean} task using Briggs weighting. For each frequency band and epoch, we produced Stokes~$I$ images and measured the local rms noise around the target position. The source flux density was measured from the image peak intensity at the optical position of Bi'an when a radio counterpart was detected. For non-detections, we report the local rms noise and adopt a $3\sigma$ upper limit on the radio flux density.

\begin{figure}
    \centering
    \includegraphics[width=\linewidth]{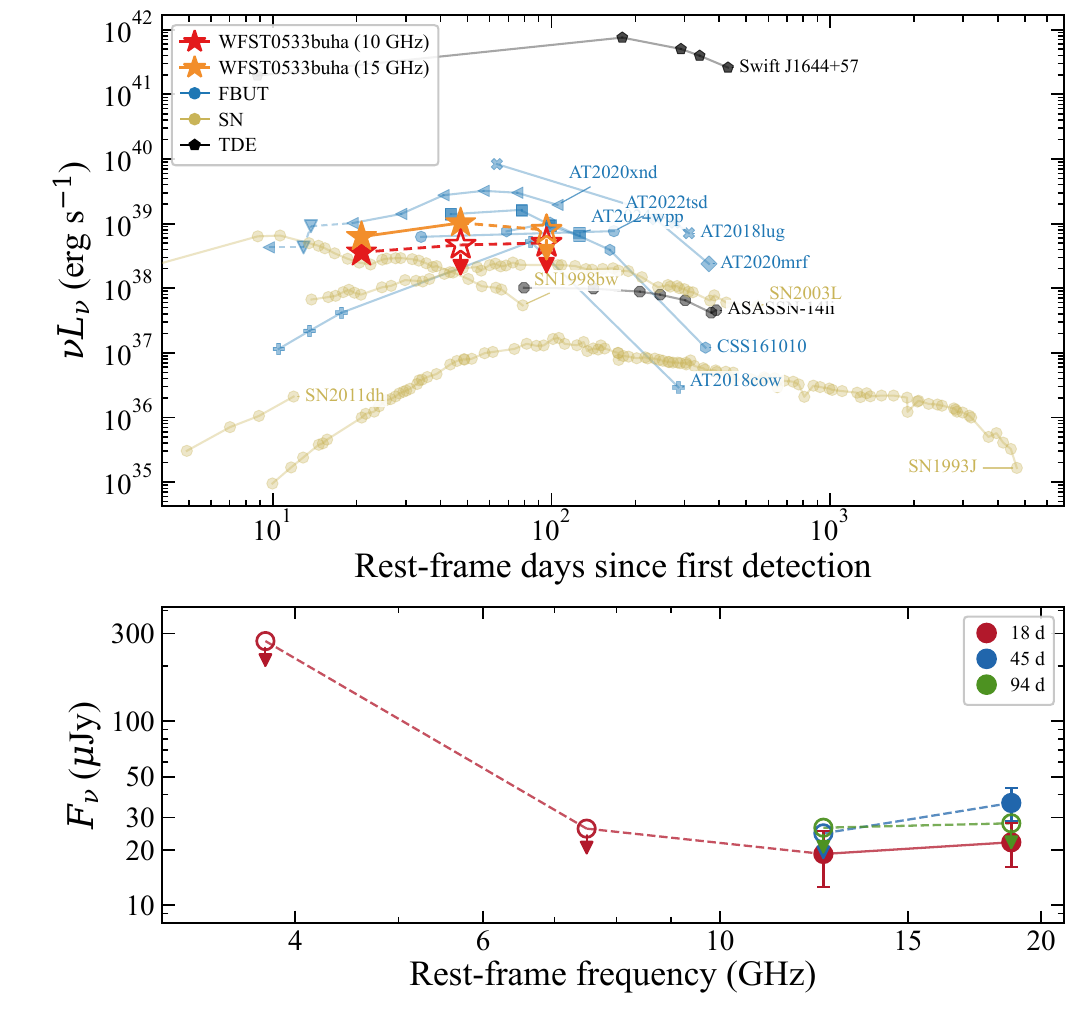}
    \caption{Radio observations of Bi'an.
    Top: radio light curve of Bi'an compared with other transients from the literature.
    Red and orange pentagrams correspond to 10 and 15 GHz measurements, respectively; open pentagrams with downward arrows denote $3\sigma$ upper limits.
    The phases are relative to the WFST discovery date in the rest frame. For Bi'an, a fiducial redshift of \(z=0.25\) is adopted to calculate the radio luminosities and rest-frame frequencies. For the comparison sources, radio bands closest to 10 GHz are adopted. The comparison sample includes FBUTs \citep{Ho_2019,Margutti_2019,Ho_2020,Coppejans_2020,Yao_2022,Ho_2023a,Nayana_2025}, radio-detected supernovae (SNe) \citep{Kulkarni_1998,Weiler_2007,Soderberg_2005,Horesh_2013}, and TDEs \citep{Zauderer2011,Berger2012,Zauderer2013,Eftekhari2018,Alexander2016}.
    Bottom: multi-epoch radio SEDs of Bi'an.
    Solid circles and open ones with downward arrows denote detections and $3\sigma$ upper limits, respectively.}
    \label{fig:radio_lc}
\end{figure}

A weak radio source was discovered close to the optical position of Bi'an in several X- and Ku-band images (Figure \ref{fig:radio_lc}). On 2025 March 29, the measured peak flux densities at 10 and 15 GHz were 19 and 22 $\mu$Jy beam$^{-1}$, corresponding to signal-to-noise ratios of approximately $3.0\sigma$ and $3.7\sigma$, respectively. On May 1, the 15 GHz measurement was 36 $\mu$Jy beam$^{-1}$ ($4.9\sigma$), while no significant emission was detected at 10 GHz. Neither band yielded a significant detection on July 1. To further assess the reliability of the radio detection, we jointly imaged the March 29 and May 1 observations. The combined 15 GHz Briggs-weighted image gives $24.8\pm4.7~\mu$Jy beam$^{-1}$ at the optical position, with a signal-to-noise ratio of 5.46 relative to the local image rms. Natural weighting gives a consistent $5.36\sigma$ result. The fitted radio centroid is offset by $0.47\arcsec$ from the optical position, supporting an association with Bi'an. At 10 GHz, the combined images yield only a marginal $2.3$--$2.4\sigma$ positive signal. The joint 15 GHz radio image and its contours overlaid on the archival HSC image are presented in Figure~\ref{fig:radio_position} (Appendix A).

\begin{table*}
  \centering
  \begin{minipage}{0.8\textwidth}
    \centering
    \caption{Summary of VLA observations of Bi'an}
    \label{tab:obs_summary}

    \begin{tabular*}{\linewidth}{@{\extracolsep{\fill}}lcccccrc}
      \hline\hline
      Date (UT) & Array Config. & $\nu$ & $b_\mathrm{maj}$ & $b_\mathrm{min}$ &
      PA & RMS & S$_{\rm peak}$ \\
      & & (GHz) & (arcsec) & (arcsec) & ($^\circ$) &
      ($\mu$Jy beam$^{-1}$) & ($\mu$Jy beam$^{-1}$) \\
      \hline
      2025-03-29 & D & 3  & 23  & 18  & -8.6 & 91  & \ldots \\
      2025-03-29 & D & 6  & 11  & 9.3 & 3.6  & 8.7 & \ldots \\
      2025-03-29 & D & 10 & 7.2 & 5.9 & -6.3 & 6.4 & 19 \\
      2025-03-29 & D & 15 & 4.9 & 4.0 & -12  & 5.9 & 22 \\
      2025-05-01 & D & 10 & 7.4 & 5.9 & -10  & 8.2 & \ldots \\
      2025-05-01 & D & 15 & 5.1 & 4.0 & -19  & 7.3 & 36 \\
      2025-07-01 & C & 10 & 2.2 & 1.8 & -2.4 & 8.8 & \ldots \\
      2025-07-01 & C & 15 & 1.5 & 1.2 & -1.7 & 9.3 & \ldots \\
      \hline
      Stacked & D & 15 & 4.91 & 4.06 & \ldots & 4.55 & $24.83\pm4.71$ \\
      \hline
    \end{tabular*}
    \tablecomments{The stacked row combines the 2025-03-29 and 2025-05-01 observations with Briggs weighting ($\mathrm{robust}=0.5$). Its flux uncertainty includes a 5\% calibration term added in quadrature to the image rms; the detection significance is $5.46\sigma$. }
  \end{minipage}
\end{table*}

\subsection{X-Ray Observations} \label{subsec:xray}
\begin{table*}
  \centering
  \begin{minipage}{0.8\textwidth}
    \centering
    \caption{Summary of X-ray follow-up observations of Bi'an}
    \label{tab:X-ray}

    \begin{tabular*}{\linewidth}{
      @{\extracolsep{\fill}}lcccc
    }
      \hline\hline
      Avg. MJD &
      Phase\textsuperscript{a} &
      Exposure time\textsuperscript{b} &
      Count-rate limit\textsuperscript{c} &
      Flux limit\textsuperscript{c} \\
      & (day) & (s) & (cts s$^{-1}$) &
      (\si{erg\,cm^{-2}\,s^{-1}}) \\
      \hline
      60743.38 & $5.63\pm0.01$  & 4743  & \num{1.7e-3} & \num{3.3e-14} \\
      60748.40 & $10.65\pm0.4$   & 7219  & \num{1.5e-3} & \num{3.0e-14} \\
      60749.76 & $12.01\pm0.05$ & 12262 & \num{1.0e-3} & \num{2.0e-14} \\
      60786.40 & $48.65\pm0.2$   & 16760 & \num{7.3e-4} & \num{1.4e-14} \\
      \hline
      \multicolumn{5}{c}{Stacked second and third observations} \\
      \hline
      60749.00 & $11.25\pm0.9$ & 19480 & \num{7.9e-4} & \num{1.5e-14} \\
      \hline
    \end{tabular*}

    \tablecomments{
    \textsuperscript{a} Phases are measured relative to the first detection at MJD 60737.748; uncertainties are quoted at the $1\sigma$ confidence
    level. \textsuperscript{b} Exposure times are the sums of the FXT-A and
    FXT-B exposures. \textsuperscript{c} Count-rate and flux upper limits are
    quoted at the $3\sigma$ confidence level. Flux limits refer to the
    rest-frame 0.5--10 keV band at $z=0.25$.
    }
  \end{minipage}
\end{table*}
To explore the X-ray emission from Bi'an, X-ray follow-up observations were triggered following its discovery using the Follow-up X-ray Telescope (FXT) on board the Einstein Probe \citep[EP;][]{Yuan2022,FXT}. All FXT observations were processed uniformly using the standard \texttt{fxtchain} pipeline in the FXT Data Analysis Software, which performs event calibration and screening \citep{Jia2025}. Subsequent analyses were based on the resulting Level 2 event files. The initial follow-up observation was conducted approximately 5.6 days after the first detection. However, with an exposure time of 4743 s, no significant X-ray source was detected within the FXT error circle at the location of Bi'an. Additionally, no significant detection was found in the three subsequent follow-up observations.

The details of the four observations and the resulting upper limits are summarized in Table \ref{tab:X-ray}. Count rate limits are derived using the \texttt{XIMAGE} package\footnote{\url{https://heasarc.gsfc.nasa.gov/docs/rosat/faqs/data_src_faq3.html}}, which is distributed and maintained by \cite{Heasoft}. To convert the count rate limits to flux limits, the \texttt{modelflux} tool is employed, as part of the Chandra Interactive Analysis of Observations (CIAO) software package \citep{CIAO}. The model is assumed to be \texttt{phabs$\times$powerlaw}, with the Galactic hydrogen column density fixed at $n_{\text{H}} = 3.54 \times 10^{20}$~cm$^{-2}$, and a power-law index of 1.5 (similar to the case of AT~2018cow \citep{Margutti_2019}). Intrinsic absorption is not considered in this analysis. The flux limits from 0.5 to 10~keV in the rest frame are derived after applying a K-correction at a redshift of $z=0.25$.

\begin{figure}[htbp]
    \centering
    \includegraphics[width=\columnwidth]{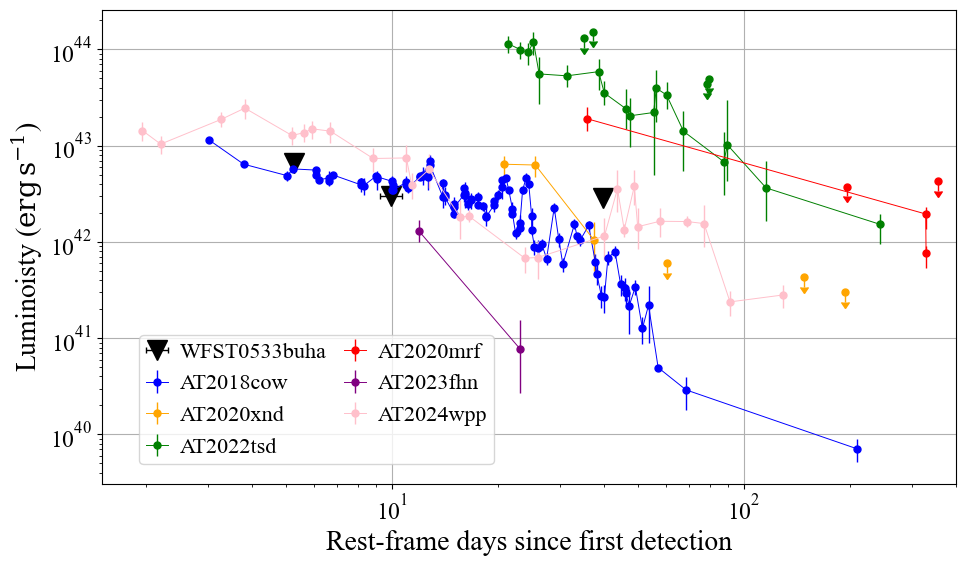}
    \caption{X-ray luminosity upper limits of Bi'an in the rest-frame 0.5–10 keV band, assuming a fiducial redshift of z = 0.25. X-ray light curves of other FBUTs including AT~2018cow \citep{Margutti_2019}, AT~2020xnd \citep{Bright_2022}, AT~2022tsd \citep{Ho_2023a}, AT~2020mrf \citep{Yao_2022} and AT~2023fhn \citep{Chrimes_2024a} are shown for comparison.}
    \label{fig:X-ray}
\end{figure}

The upper limits of the X-ray luminosity of Bi'an and other known FBUTS \citep{Margutti_2019,Yao_2022,Bright_2022,Ho_2023a,Chrimes_2024a} are shown in Figure \ref{fig:X-ray}.

\section{Observational Properties} \label{sec:an}
Bi'an shows the rapid optical evolution and blue, predominantly featureless spectra characteristic of FBUTs. Although some cataclysmic variables (CVs) and rapidly evolving SNe can share these optical properties, the combined photometric, spectroscopic, and radio observations favor an FBUT interpretation. In particular, the radio luminosity and detection epochs disfavor a conventional CV origin, while the faster optical evolution and absence of identifiable He I features distinguish Bi'an from the SNe Ibn considered here. Detailed comparisons supporting its classification as an FBUT candidate are presented in Appendix~B.

\subsection{Host Galaxy} \label{subsec:host}
\begin{figure}[ht]
    \centering
    \includegraphics[width=0.8\columnwidth]{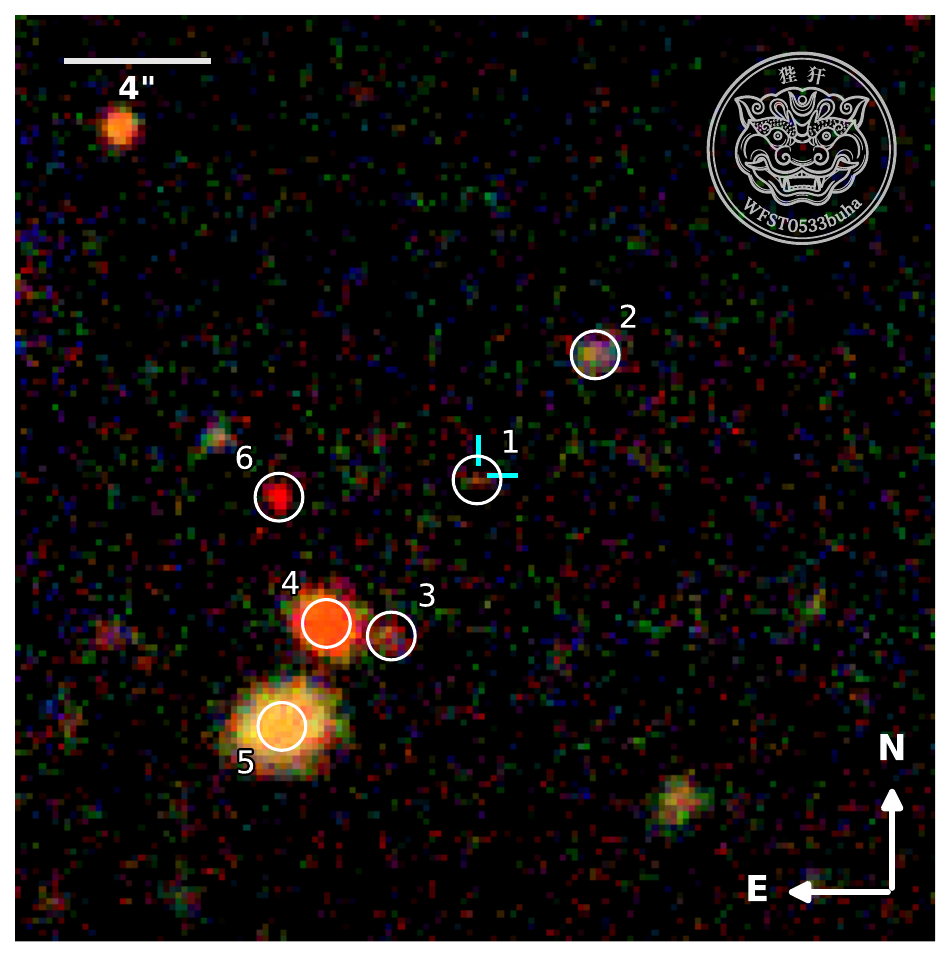}
    \caption{HSC-SSP cutout around Bi'an \citep{Aihara_2022}. The cyan cross marks the transient position and the numbered circles correspond to host candidates listed in Table~\ref{tab:buha_host_candidates}. An emblem designed for Bi'an appears in the upper-right corner.}
    \label{fig:host}
\end{figure}

The HSC PDR3 images reveal an extremely faint source coincident with Bi'an (candidate 1 in Figure~\ref{fig:host}) and five nearby host candidates at angular separations of $4.7$--$8.6$ arcsec. The latter show no detectable extended structure reaching the transient position. Their projected offsets span approximately $32$--$53$ kpc over their respective photometric-redshift intervals, corresponding to approximately $11$--$19$ half-light radii for the resolved candidates (Table~\ref{tab:buha_host_candidates}). Such extremely large offsets would be unusual for an FBUT associated with a young stellar population, although an association cannot be completely excluded. The faint source coincident with the transient position therefore remains the more plausible host candidate. Because of the faintness, no additional spectrum of the putative host galaxy was obtained. 

For $0.05\leq z\leq0.30$, the host candidate would have $M_i\approx-10.7$ to $-14.9$ mag, neglecting K-corrections, consistent with a very faint dwarf galaxy at the low-luminosity end of the known FBUT host population \citep{Nugent_2026}.

\begin{table*}[t]
\centering
\caption{HSC host-galaxy candidates near Bi'an.}
\label{tab:buha_host_candidates}
\begingroup
\scriptsize
\renewcommand{\arraystretch}{1.1}
\setlength{\tabcolsep}{2pt}
\begin{tabular*}{\textwidth}{@{\extracolsep{\fill}}cllccccccccc@{}}
\hline\hline
ID & R.A. (J2000) & Dec. (J2000) & \multicolumn{3}{c}{Offset} & Photo-$z$ & $g$ & $r$ & $i$ & $z$ & $y$ \\
 & & & (arcsec) & (kpc) & ($R_e$) & & \multicolumn{5}{c}{(AB mag)} \\
\hline
1 & 10:44:34.565 & +04:28:12.54 & 0.09 & $\cdots$ & $\cdots$ & $\cdots$ & $>26.59$ & $>26.88$ & $26.02\pm0.20$ & $>25.65$ & $>24.58$ \\
2 & 10:44:34.350 & +04:28:15.95 & 4.67 & 34.4--39.5 & 13.3 & $1.06^{+0.49}_{-0.30}$ & $24.89\pm0.10$ & $24.94\pm0.09$ & $24.58\pm0.08$ & $24.17\pm0.11$ & $>24.09$ \\
3 & 10:44:34.720 & +04:28:08.29 & 4.88 & 32.4--37.8 & 15.7 & $0.71^{+0.17}_{-0.12}$ & $25.83\pm0.32$ & $25.63\pm0.20$ & $24.83\pm0.10$ & $25.00\pm0.34$ & $>24.16$ \\
4 & 10:44:34.839 & +04:28:08.64 & 5.67 & 38.2--40.5 & 18.6 & $0.65^{+0.05}_{-0.04}$ & $24.86\pm0.15$ & $23.26\pm0.02$ & $22.14\pm0.01$ & $21.83\pm0.02$ & $21.48\pm0.03$ \\
5 & 10:44:34.919 & +04:28:05.83 & 8.57 & 49.4--52.9 & 10.8 & $0.48^{+0.03}_{-0.03}$ & $23.15\pm0.03$ & $22.28\pm0.01$ & $21.94\pm0.01$ & $21.81\pm0.03$ & $21.62\pm0.05$ \\
6 & 10:44:34.925 & +04:28:12.07 & 5.36 & 36.8--45.4 & $\cdots$ & $1.15^{+2.96}_{-0.20}$ & $>26.57$ & $>26.91$ & $24.90\pm0.07$ & $23.63\pm0.06$ & $23.06\pm0.09$ \\
\hline\hline
\end{tabular*}
\par\vspace{2pt}
\parbox{\textwidth}{\footnotesize
\textit{Note.} Magnitudes are HSC forced cModel AB measurements; limits are $3\sigma$. Photo-$z$ errors and physical-offset ranges use the DEmP 68\% intervals. $R_e$ denotes the circularized $i$-band half-light radius, unresolved for candidates 1 and 6. Candidates 3--5 share a deblending parent.}
\endgroup
\end{table*}



\subsection{Optical Light Curve} \label{subsec:olc}
Bi'an shows a very rapid optical rise. The first \textit{u}-band detection of $22.08 \pm 0.11$ mag was derived from coadded images at MJD 60737.748, and the source brightened to $19.63 \pm 0.01$ mag at MJD $60740.743$, corresponding to an increase of $\sim 2.5$ mag in 3 days in the observer frame. The extinction-corrected apparent peak magnitudes are $19.63 \pm 0.01$, $19.70 \pm 0.01$, and $20.10 \pm 0.01$ in \textit{ugr} bands, corresponding to absolute magnitudes of $-20.68$, $-20.62$, and $-20.21$ at the redshift of 0.25, respectively. The deep, early $u$-band detection provides a rare view of the rapid rise and reveals the extremely blue color of the transient at early times.

The decline speed of Bi'an is also rapid. The object becomes more than 2-mag fainter than the maximum in 9 days after the peak. We model the near-peak $g$-band light curve using Gaussian-process regression with a squared-exponential kernel, incorporating pre-discovery forced-flux measurements. The resulting posterior distributions yield observer-frame half-maximum rise and decline times of $2.20^{+0.10}_{-0.09}$ and $3.39^{+0.21}_{-0.20}$ days, respectively, giving a total duration of $5.60\pm0.20$ days. The uncertainties are conditional on the optimized kernel hyperparameters. We compared the light curve of Bi'an with AT~2018cow, AT~2020xnd and AT~2024wpp in Figure \ref{fig:lc}. Its post-maximum decline is broadly similar to those of AT~2018cow and AT~2020xnd and is substantially faster than that of AT~2024wpp.

\begin{figure}
	\plotone{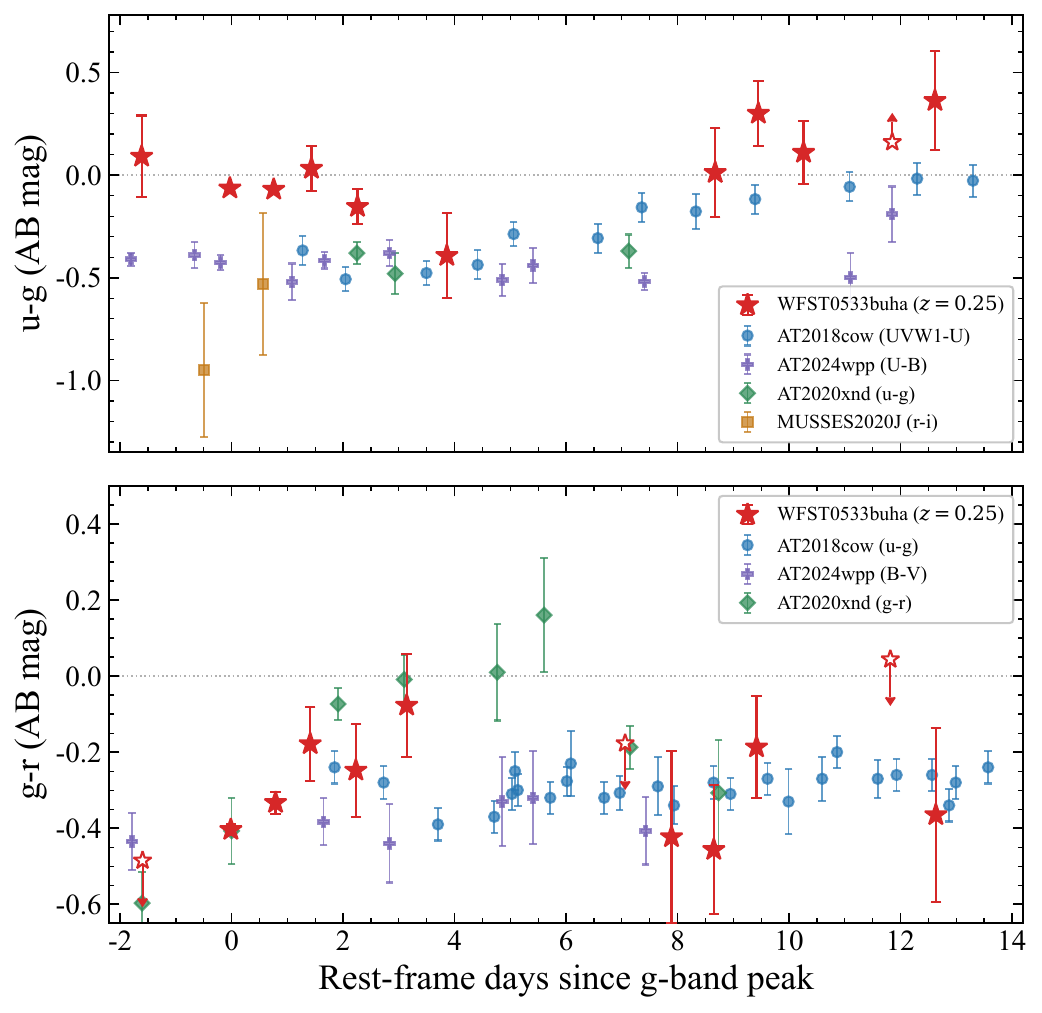}
	\caption{Color evolution of Bi'an compared with AT~2018cow ($z=0.014$; \citealt{perley_2019}), AT~2024wpp ($z=0.087$; \citealt{LeBaron_2026}), AT~2020xnd ($z=0.243$; \citealt{Perley2021}) and MUSSES2020J ($z=1.063$; \citealt{jiang_2022}) at comparable rest-frame wavelengths. Open symbols with arrows denote sources for which the $r$-band detection limit sets the reddest possible $g-r$ color. Fiducial redshift $z = 0.25$ for Bi'an was adopted.}
	\label{fig:color}
\end{figure}

Adopting a fiducial redshift $z=0.25$ and correcting for Galactic extinction, we present the color evolution of Bi'an in Figure \ref{fig:color}. Among the FBUTs reported in the literature, Bi'an is the third event, after AT~2018cow and AT~2024wpp, with extensive $u$-band coverage. Given the large redshift scatter among current FBUT sample, we compare colors with similar rest-frame wavelength coverage: the observed $u-g$ color of Bi'an is compared with $Swift$/UVOT $UVW1-U$, $U-B$, $u-g$ and $r-i$ for AT~2018cow, AT~2024wpp, AT~2020xnd and MUSSES2020J, respectively, while its $g-r$ color is compared with $u-g$, $B-V$, and $g-r$ for AT~2018cow, AT~2024wpp and AT~2020xnd.

Before and around the $g$-band maximum, Bi'an has a distinctly redder observer-frame $u-g$ than AT~2018cow and AT~2024wpp, reaching $u-g=-0.064\pm0.013$ mag near maximum. Its $u-g$ color subsequently becomes bluer but remains redder than that of AT~2018cow, and becomes redder again at approximately 9--10 rest-frame days. In contrast, the $g-r$ color of Bi'an remains blue and evolves broadly consistently with those of AT~2018cow and AT~2024wpp. The color difference confined to the NUV wavelengths may suggest a difference in the underlying spectral energy distribution, possibly due to additional suppression at UV/NUV wavelengths. The implications for possible UV diversity among FBUTs are discussed in Section~\ref{uv}.

AT~2022tsd showed luminous optical flares on minute timescales, revealing a form of late-time activity not previously seen in FBUTs \citep{Ho_2023a}. A recent search for similar flares in previous FBUTs found no robust minute-timescale events outside AT~2022tsd \citep{Jayaraman_2026}. We also searched for short-timescale optical variability on 2026 February 11, about 342 days after the $g$-band peak in the observer frame, using 56 consecutive 30s WFST exposures in the $w$ band, providing continuous coverage over approximately 30 minutes. The $w$-band exposures reached a median $5\sigma$ limiting magnitude of $\sim23.7$ mag, but no source was detected at the position of Bi'an (Figure \ref{fig:lc}). In addition, WFST monitored the field with a daily cadence during the seasonal visibility window but did not yield additional detections. These observations suggest there is no clear evidence for either minute-timescale optical flares or luminous activities on day timescales.

\subsection{Spectra}

Figure~\ref{fig:spec} compares the two spectra of Bi'an, obtained approximately 2 and 11 observer-frame days after maximum, with those of AT~2018cow, AT~2020xnd, SN Ibn~2019deh, and SN IIb~2018gjx. At the illustrative redshift of $z=0.25$, the predominantly blue continua broadly resemble those of AT~2018cow and AT~2020xnd. Neither spectrum shows convincing counterparts to the prominent features in the comparison spectra of SN~2019deh and SN~2018gjx.

The first spectrum shows a weak, broad emission-like feature near 5125 \AA\ in the observer frame. A Gaussian fit with a local quadratic continuum, fixing the center at 5125 \AA\ and the FWHM at 200 \AA, yields a formal amplitude significance of $5.1\sigma$. The second spectrum is noisier and shows no convincing broad features after binning to 20 \AA.

\subsection{Multiwavelength Properties}

At the illustrative redshift of $z=0.25$, the EP upper limits rule out X-ray luminosities comparable to those of AT~2020mrf and AT~2022tsd at similar phases, while allowing an AT~2018cow-like luminosity (Figure~\ref{fig:X-ray}; \citealt{Margutti_2019,Bright_2022,Yao_2022,Ho_2023a,Chrimes_2024a}). The redshift dependence of these constraints is discussed in Section~\ref{subsec:redshift}.

The radio emission may arise from an outflow interacting with the surrounding medium. However, the limited frequency coverage does not constrain the synchrotron peak, preventing reliable estimates of the shock properties. We therefore do not attempt detailed modeling of the radio emission.

\subsection{Blackbody Fits}
\begin{figure}
    \centering
    \includegraphics[width=0.8\linewidth]{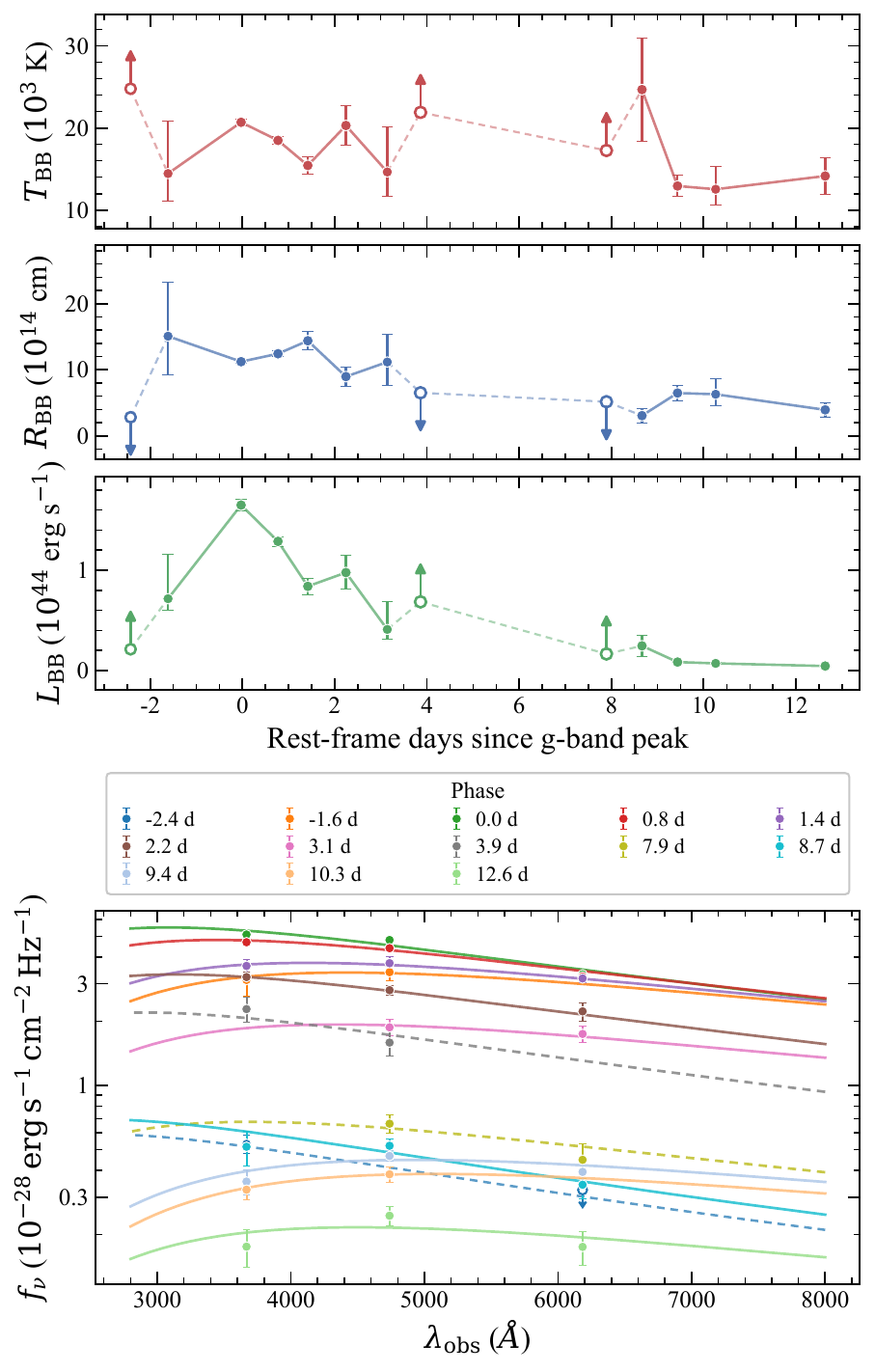}
    \caption{Blackbody modeling of the $ugr$ photometry of Bi'an assuming a fiducial redshift of $z=0.25$. The upper panels show the evolution of the blackbody temperature, radius, and bolometric luminosity. The bottom panel shows the observed SEDs and the best-fitting blackbody models. Dashed curves indicate cases where the blackbody parameters are constrained by the boundaries of the allowed parameter space, either because only one photometric detection and an upper limit are available or because the fit reaches the adopted parameter-search bounds.}
    \label{fig:bb}
\end{figure}

We fit single-temperature blackbody models to the $ugr$ photometry at the fiducial redshift of $z=0.25$ (Figure \ref{fig:bb}). The data were corrected for Galactic extinction as described in Section \ref{subsec:olc}, with no correction for host extinction. Model fluxes were integrated through the WFST bandpasses, and the temperature and radius were estimated using \texttt{emcee} \citep{ForemanMackey_2013} with a Gaussian likelihood based solely on the photometric errors. We report posterior medians and 68\% credible intervals. For the seven epochs with $ugr$ photometry, the minimum reduced $\chi^2$ ranges from 0.032 to 92.3, with one degree of freedom per epoch. The two epochs near peak have reduced $\chi^2$ values of 92.3 and 14.1, indicating that a single blackbody does not fully describe their SEDs.

Near optical peak, the fits yield $T_{\rm BB}=(2.07\pm0.04)\times10^4$ K and $R_{\rm BB}=(1.12\pm0.02)\times10^{15}$ cm, corresponding to $L_{\rm BB}=1.65^{+0.06}_{-0.05}\times10^{44}$ erg s$^{-1}$. The inferred temperature remains above $10^4$ K during the first $\sim10$ rest-frame days. The quoted uncertainties are statistical and conditional on
the adopted single-temperature blackbody model.

We also fit blackbody continua to the Galactic-extinction-corrected spectra over 4000--7200 \AA\ in the observer frame, using robust least squares and including an additional 5\% flux uncertainty. Both spectra and their uncertainties were corrected for Galactic extinction using the law of \citet{Cardelli_1989}, adopting $E(B-V)=0.04$ mag and $R_V=3.1$. No correction for host extinction was applied. The resulting temperatures are $(2.34\pm0.18)\times10^4$ K and $(2.30\pm0.19)\times10^4$ K, with reduced $\chi^2$ values of 1.68 and 1.28, respectively, each for 52 nominal degrees of freedom. For comparison, we also performed blackbody fits using only the $g$- and $r$-band photometry. Around peak, the $gr$-only fit yields
$T_{\rm BB}=3.15^{+0.23}_{-0.20}\times10^4$ K,
compared with $(2.07\pm0.04)\times10^4$ K from the $ugr$ fit. Extrapolating the $gr$-based blackbody model to the $u$ band predicts a $u$-band flux approximately 0.24 mag brighter than observed, indicating a flux deficit at shorter wavelengths relative to the optical blackbody continuum.

\section{Discussion} \label{sec:disc}

\subsection{Effects of Redshift Uncertainty}\label{subsec:redshift}

Since Bi'an has no secure redshift measurement, we examine its inferred properties over $0.05\leq z\leq0.30$, with their redshift dependence shown in Figure~\ref{fig:redshift_properties} in the Appendix.

Over $z=0.20$--0.30, the rest-frame colors inferred under the blackbody approximation depend only weakly on the assumed redshift. Synthetic photometry of the fitted blackbody SEDs through fixed rest-frame WFST bandpasses gives differences of approximately 0.02--0.05 mag in $u-g$ and 0.02--0.04 mag in $g-r$ between $z=0.20$ and $z=0.30$. These small differences do not affect our conclusion regarding possible diversity in the UV properties of FBUTs.

Over the full range $0.05\leq z\leq0.30$, the near-peak blackbody temperature varies from approximately $1.7\times10^4$ to $2.2\times10^4$ K, while the radius and bolometric luminosity span approximately $(0.28$--$1.28)\times10^{15}$ cm and $5.1\times10^{42}$--$2.5\times10^{44}$ erg s$^{-1}$, respectively. The evidence for a hot continuum is therefore robust, whereas the inferred photospheric size and luminosity depend strongly on distance. These quantities remain approximate because a single blackbody does not fully describe the near-peak SED.

The interpretation of the X-ray limits also depends on distance. Over $z=0.20$--$0.30$, X-ray emission comparable to that observed in AT~2022tsd or AT~2020mrf would remain inconsistent with the EP observations, whereas the comparison with AT~2018cow depends more strongly on the adopted distance and phase. The radio detection and its positional association with Bi'an are independent of the assumed redshift, whereas the inferred radio luminosity depends on distance.

\subsection{UV Diversity among FBUTs}\label{uv}

The UV behavior of FBUTs remains poorly constrained because few events have sufficiently early and densely sampled UV light curves. Bi'an provides a valuable comparison with AT~2018cow because the $u-g$ color at \(z\sim0.25\) is comparable to $UVW1-U$ at the redshift of AT~2018cow (i.e., \(z\sim0.014\)). Although the optical $g-r$ evolution of Bi'an is broadly consistent with that of AT~2018cow, Bi'an is systematically redder in $u-g$. The difference is therefore concentrated at shorter rest-frame wavelengths. Because the comparison involves different filter transmission curves and an assumed redshift for Bi'an, the observed color differences should be interpreted as suggestive rather than conclusive evidence for intrinsic UV diversity.

If this difference reflects intrinsic diversity among FBUTs, several physical explanations are possible. One possible explanation is enhanced line blanketing or bound-free opacity in UV wavelengths, which would suppress the observed $u$-band flux while leaving $g-r$ comparatively unchanged. This possibility also illustrates an important limitation of blackbody fits based only on optical photometry. An optical blackbody temperature does not necessarily predict the ultraviolet spectral energy distribution \citep{perley_2019,LeBaron_2026}. FBUTs with similar optical temperatures may indeed show substantial differences in the UV opacity and ionization state. In this picture, one possible interpretation is that the photosphere of Bi'an has a lower ionization state and stronger UV line/bound-free opacity than those of AT~2018cow. Such diversity could also affect the escape or reprocessing of high-energy radiation and may contribute to the X-ray diversity of the class. Alternatively, attenuation by circumstellar dust prior to its destruction could also suppress the short-wavelength emission \citep{Metzge_Perley_2023}; however, the blueward evolution occurs after the peak and the lack of near-infrared observations prevents us from testing the predicted dust-echo signature.

Although the current observations do not uniquely identify the physical origin of Bi'an, its rapid optical evolution, high inferred luminosity, and hot continuum are broadly consistent with scenarios involving a relatively low ejecta mass and an additional power source beyond radioactive decay. The spatially coincident radio emission may be associated with an outflow interacting with the surrounding medium, while the X-ray non-detections leave open the possibilities of a weaker central engine, absorption, or efficient reprocessing of high-energy radiation \citep{Margutti_2019,Metzger_2022}.

An unbiased, systematically selected sample is thus needed to test the above interpretation. We note that the WFST DH\textit{ugr} can identify FBUTs at \(z\sim0.2\)--0.3 and measure their early \(u-g\) evolution, enabling direct comparisons with the \(\mathrm{UVW1}-U\) evolution of nearby events such as AT~2018cow. Joint X-ray, radio, and spectroscopic follow-ups will be essential for determining whether FBUTs form a single blackbody-like population or show genuine diversity in their UV opacity and ionization conditions.

With the advent of new-generation deep, high-cadence time-domain surveys such as WFST DH$ugr$, the samples of both FBOTs and FBUTs are expected to grow rapidly. In particular, the combination of wide-field coverage, deep imaging, and high-cadence $u$-band observations makes WFST particularly well suited to discovering energetic, rapidly evolving transients at very early phases. WFST has already accumulated dozens of high-quality fast blue transients with early multi-band coverage, providing new opportunities to investigate the physical origins and diversity of these events (Jia et al., 2026, in prep). 

\section{Conclusions}

In this Letter, we report the discovery and follow-up observations of the fast blue ultraluminous transient candidate WFST0533buha (Bi'an) by the WFST Deep High-cadence $ugr$-band Survey. The transient shows rapid optical evolution, with a light curve broadly resembling those of AT~2018cow and AT~2020xnd. Its extensive $u$-band coverage reveals a systematically redder $u-g$ evolution than the approximately wavelength-matched colors of AT~2018cow and AT~2024wpp, suggesting possible diversity in the short-wavelength properties of FBUTs.

The multiwavelength observations provide additional constraints on the nature of Bi'an. The spatially coincident radio emission may be associated with an outflow interacting with the surrounding medium, while the X-ray non-detections do not rule out a weaker or obscured central engine. Although its rapid optical evolution and high inferred luminosity suggest a relatively low ejecta mass and an additional power source, the current observations do not uniquely distinguish among the proposed progenitor scenarios.

The early $u$-band detection of Bi'an demonstrates the value of deep, high-cadence blue-band observations for identifying and characterizing FBUTs at early phases. The ongoing WFST DH$ugr$ survey is expected to build a sample of FBUTs with early multiband coverage in the coming years, enabling systematic comparisons of their rise behavior, color evolution, UV properties, and multiwavelength emission. Such observations will be essential for establishing the extent of diversity in the ultraviolet and high-energy properties of FBUTs and, ultimately, for constraining the physical origins of this population.

\section*{ACKNOWLEDGEMENTS}
This work was supported by the National Natural Science Foundation of China (Grant No. 12393811), the National Key R\&D Program of China (Grant No. 2023YFA1608100), and the Strategic Priority Research Program of the Chinese Academy of Science (Grant No. XDB0550300). J.J. acknowledges support from the Japan Society for the Promotion of Science (JSPS) KAKENHI grants JP22K14069. K.M. acknowledges support from the JSPS KAKENHI grant JP24KK0070 and 23H04894. K.M. and H.K. acknowledge support from the JSPS bilateral JPJSBP120229923. L.G. acknowledges financial support from MCIN and AEI 10.13039/501100011033 under project PID2023-151307NB-I00.\\

The Wide Field Survey Telescope (WFST) is a joint facility of the University of Science and Technology of China and Purple Mountain Observatory. This work is based on the data obtained with Einstein Probe, a space mission supported by the Strategic Priority Program on Space Science of Chinese Academy of Sciences, in collaboration with the European Space Agency, the Max-Planck-Institute for extraterrestrial Physics (Germany), and the Centre National d'E$\mathring{\text{A}}$Ltudes Spatiales (France). This research is based in part on data collected at the Subaru Telescope (S25A-024; PI: K. Maeda), which is operated by the National Astronomical Observatory of Japan. The observations by Gemini-South (S25A-079 and GS-2025A-Q-101; PI: K. Maeda) were carried out within the framework of the Subaru-Gemini time exchange program. We are honored and grateful for the opportunity of observing the Universe from Maunakea, which has cultural, historical, and natural significance in Hawaii.

\appendix

\section{Radio Imaging and Positional Association}
\label{app:radio}

\begin{figure}[!t]
    \centering
    \includegraphics[width=\linewidth]{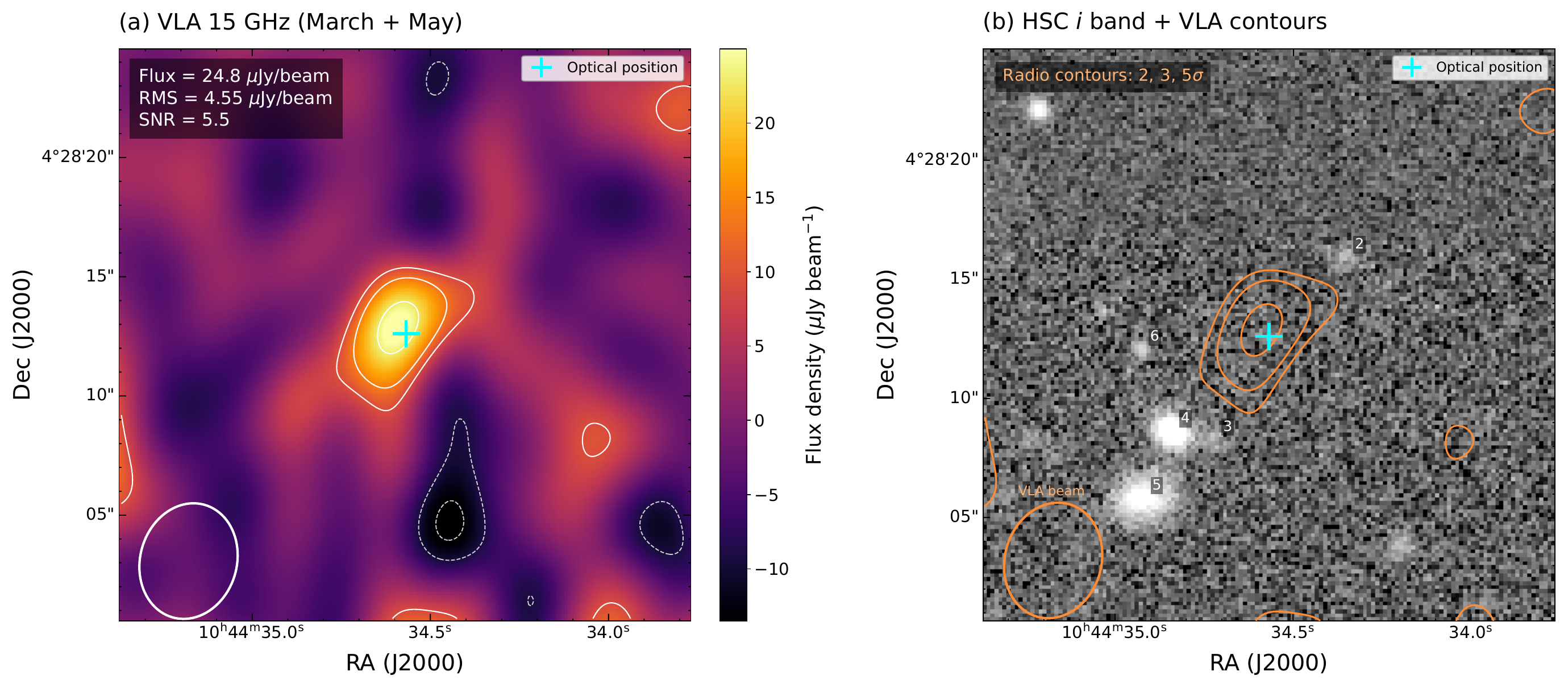}
    \caption{
    Joint 15 GHz VLA image from the 2025 March 29 and
    May 1 observations (left) and archival HSC $i$-band
    imaging of the same field with radio contours
    overlaid (right). Cyan crosses mark the optical
    position of Bi'an. Positive contours correspond
    to 2, 3, and $5\sigma$, and dashed contours indicate
    $-3$ and $-2\sigma$, where
    $\sigma=4.55~\mu$Jy beam$^{-1}$.
    The ellipses show the restoring beam of
    $4.91''\times4.06''$ at a position angle of
    $-15.9^\circ$. Numbers identify the neighboring
    host candidates listed in
    Table~\ref{tab:buha_host_candidates}.
    Each panel covers $24''\times24''$.
    }
    \label{fig:radio_position}
\end{figure}

Figure~\ref{fig:radio_position} shows the joint 15 GHz
VLA image and the radio contours overlaid on archival
HSC $i$-band imaging. The radio emission is spatially
consistent with the optical position of Bi'an and
separated from the neighboring host candidates 2--6.

\section{Assessing the Nature of Bi'an}
\label{sec:nature}

\subsection{Constraints on a Galactic Origin}

We first assess whether Bi'an could be a Galactic nova or
dwarf nova. Assuming a classical-nova peak absolute magnitude of $-10\lesssim M_{\rm peak}\lesssim-5$, the observed peak brightness would place Bi'an at a distance of approximately 0.9–9 Mpc, well beyond the Milky Way. Adopting an approximate dwarf-nova peak
magnitude range of $3\lesssim M_{g,\mathrm{peak}}\lesssim6$
mag gives a distance of 5.5--22 kpc. At these distances,
the joint 15 GHz detection corresponds to
$L_\nu\simeq(0.9$--$14)\times10^{18}\,
\mathrm{erg\,s^{-1}\,Hz^{-1}}$, approximately 3--45 times
the peak spectral luminosity of the exceptional 15.5 GHz
flare in SS Cyg. Radio emission was detected in observations
approximately 22 and 55 observer-frame days after optical
maximum, contrasting with the approximately 15-minute
duration of that flare \citep{Mooley_2017}.

Neither optical spectrum shows identifiable Balmer or helium
features. Furthermore, archival ZTF and ATLAS observations
reveal no previous outburst at $\geq5\sigma$
\citep{Masci_2023,Shingles_2021}. Together, these observations
disfavor a conventional dwarf-nova interpretation.

\subsection{Comparison with Rapidly Evolving SNe}

\begin{figure}[!htbp]
    \centering
    \includegraphics[width=\linewidth]{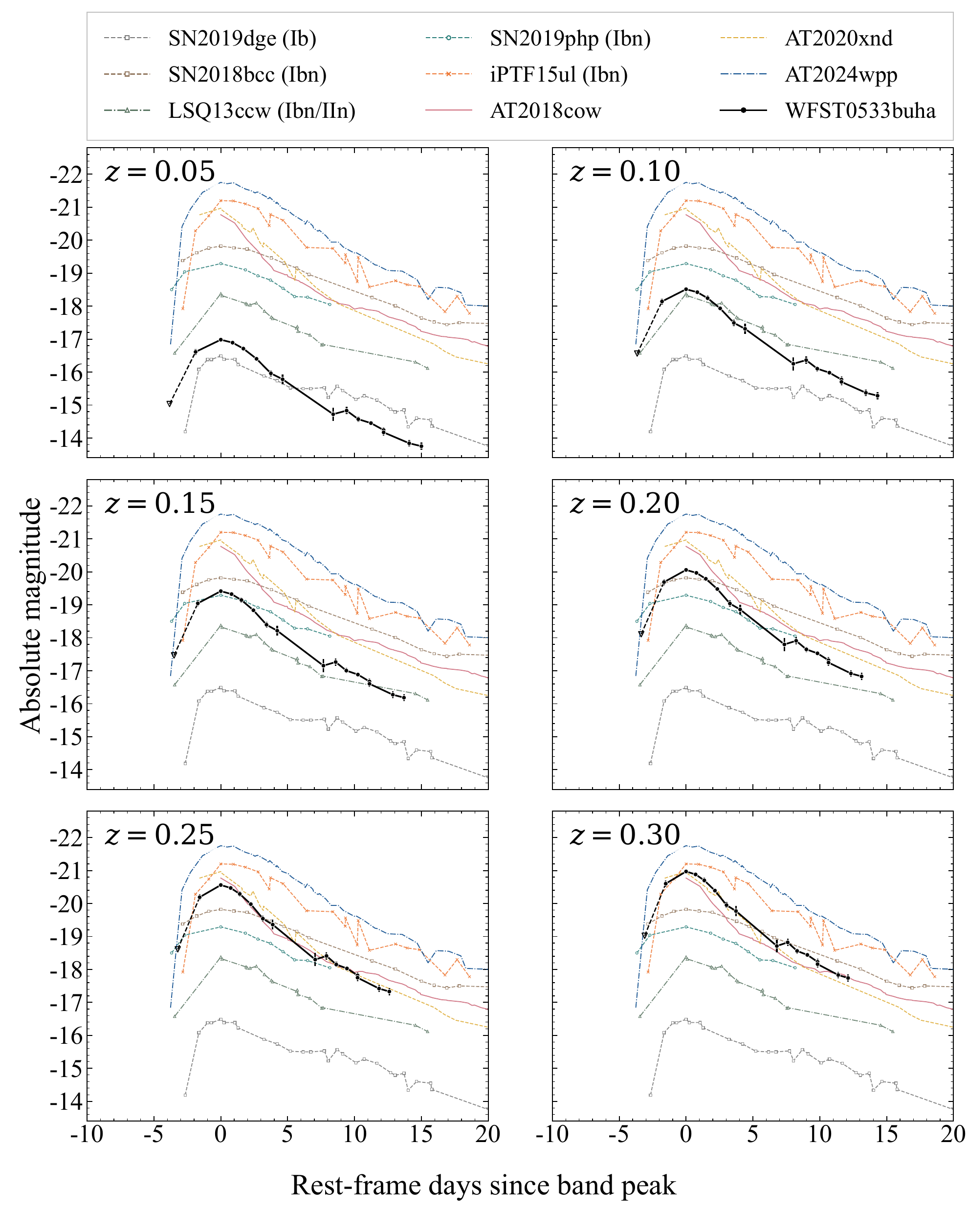}
    \caption{
    Comparison of Bi'an at the indicated trial redshifts
    with rapidly evolving SNe and representative FBUTs.
    The SN comparison data are from \citet{Ho_2023},
    \citet{Pastorello_2015}, and \citet{Hosseinzadeh_2017}.
    The FBUT light curves are the same as in
    Figure~\ref{fig:lc}. The iPTF15ul curve includes the
    published host-extinction correction, with its peak
    luminosity adopted from \citet{Ho_2020}.
    }
    \label{fig:lc_redshift}
\end{figure}

As discussed by \citet{Ho_2023}, some core-collapse
supernovae (CCSNe) can exhibit rapid photometric evolution
and blue optical colors. We therefore compare Bi'an with
rapidly evolving SNe, particularly SNe Ibn, over
a range of trial redshifts. Although its peak luminosity
overlaps with those of SNe Ibn at lower redshifts, Bi'an
evolves more rapidly, with a shorter half-maximum duration
and a steeper post-maximum decline than most Ibn events
in the comparison sample (Figure~\ref{fig:lc_redshift}).

\begin{figure}[!htbp]
    \centering
    \includegraphics[width=\linewidth]{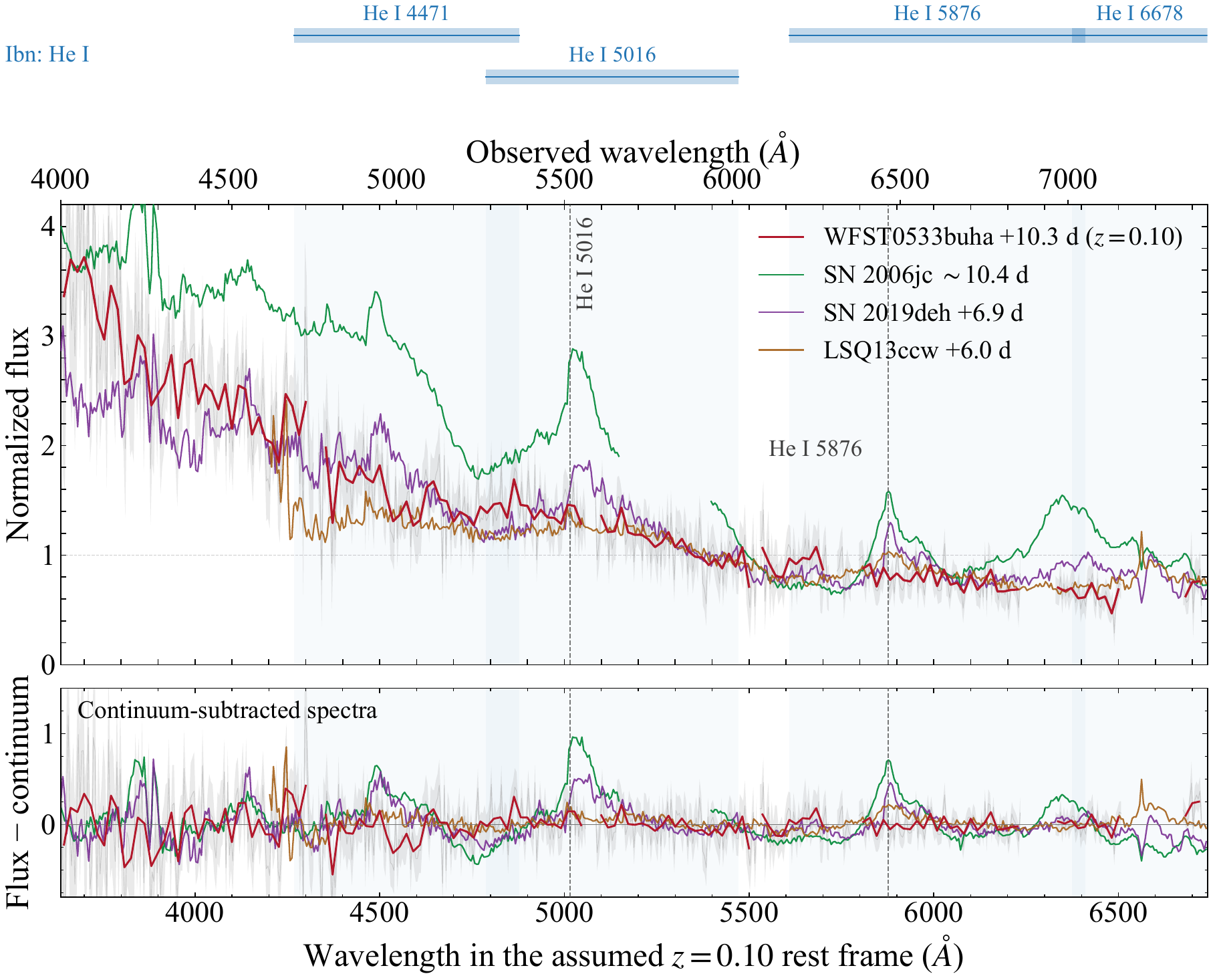}
    \caption{
    Comparison of Bi'an's second spectrum, shifted to the rest frame assuming $z=0.10$, with spectra of SNe Ibn. Labels indicate rest-frame phases relative to optical maximum \citep{Pastorello_2007,Pastorello_2015,Pellegrino_2022}. Spectra are normalized over
    5200--5600~\AA; the lower panel shows
    continuum-subtracted spectra. Bi'an is shown in
    20~\AA\ observed-frame bins in red, with its
    5~\AA\ spectrum and empirical $1\sigma$ uncertainties
    in grey. Dashed lines mark He I $\lambda5016$ and
    $\lambda5876$. Blue bars and shaded regions indicate
    the positions of selected He I transitions over
    $0.05\leq z\leq0.20$. The upper axis gives Bi'an's
    observed-frame wavelength.
    }
    \label{fig:Ibn_Icn_ranges_residual}
\end{figure}

At lower trial redshifts, several diagnostic He I
transitions fall within the observed spectral range,
but no convincing counterparts are identified in either
spectrum (Figure~\ref{fig:Ibn_Icn_ranges_residual}).
An independent template-matching analysis with SNID-SAGE
\citep{Stoppa_2026} yielded no reliable classification for
either spectrum.

The combined 15 GHz measurement, obtained by jointly imaging observations taken approximately 22 and 55 observer-frame days after optical maximum, corresponds to a spectral luminosity of $1.4\times10^{27}$--$3.8\times10^{28}\,
\mathrm{erg\,s^{-1}\,Hz^{-1}}$ over $z=0.05$--0.25. Even at the lowest trial redshift, this is about an
order of magnitude above the highest measured 15 GHz
luminosity of SN~2023fyq, recorded at 125 days after
explosion \citep{Baer-Way_2025}. This comparison provides
additional, albeit non-decisive, evidence against a
typical SN Ibn interpretation.

These comparisons disfavor both a conventional Galactic CV origin and interpretations involving the SNe Ibn considered here, supporting the FBUT candidate classification adopted in the main text. The uncertain redshift nevertheless
prevents a secure determination of Bi'an's peak luminosity.


\section{Redshift Dependence of Inferred Properties}
\label{app:redshift}

\begin{figure}
    \centering
    \includegraphics[width=\linewidth]{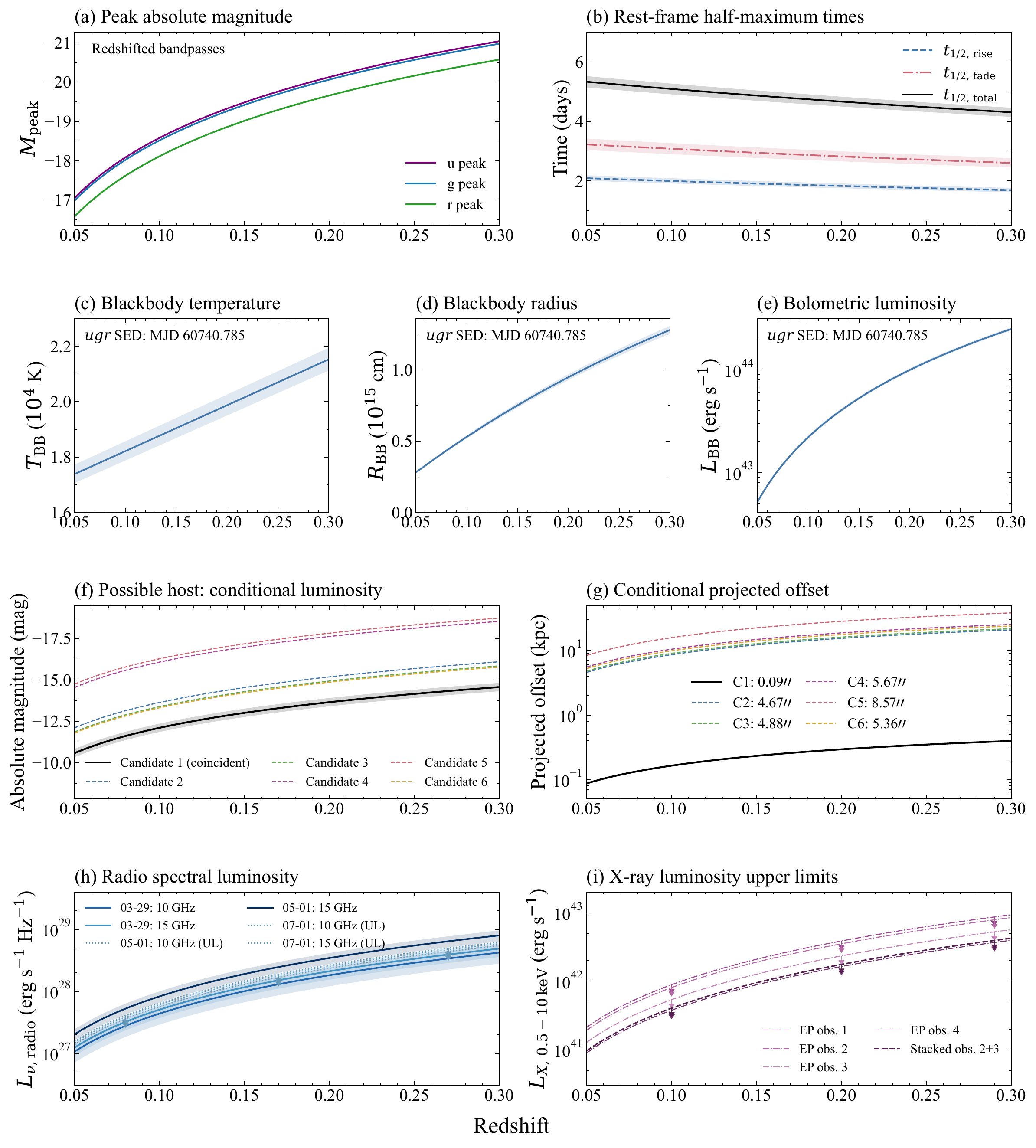}
    \caption{Dependence of the inferred properties of Bi'an on
    the assumed redshift. Panels show (a) peak absolute
    magnitudes in the shifted $ugr$ bandpasses;
    (b) rest-frame half-maximum timescales;
    (c--e) near-peak blackbody temperature, radius,
    and bolometric luminosity;
    (f--g) $i$-band absolute magnitudes and projected offsets of the host candidates, assuming each lies at the trial redshift; (h) radio spectral luminosity; and
    (i) X-ray luminosity upper limits. Shaded regions
    indicate formal uncertainties under the adopted models.}
    \label{fig:redshift_properties}
\end{figure}

Figure~\ref{fig:redshift_properties} shows the inferred
properties of Bi'an over $0.05\leq z\leq0.30$, using the
assumptions described in the main text. The observed
colors and observer-frame timescales are independent
of the adopted redshift, whereas the rest-frame
wavelengths, timescales, and inferred luminosities
depend on it.

Under the blackbody approximation, synthetic photometry using the WFST filter transmission curves in the rest frame gives color differences of approximately 0.02--0.05 mag in $u-g$ and 0.02--0.04 mag in $g-r$ between $z=0.20$ and
$z=0.30$. These differences suggest that the inferred
rest-frame colors are relatively insensitive to modest
redshift changes within this interval. At substantially
lower redshifts, the observed bands sample longer
rest-frame wavelengths, so the interpretation of the
color differences as possible UV diversity remains
dependent on the assumed redshift.

\clearpage

\bibliographystyle{aasjournalv7}
\bibliography{ref}

\end{document}